\documentclass[twocolumn,showpacs]{revtex4}

\usepackage{graphicx}
\usepackage{sidecap}

\newcommand{\be}{\begin{equation}}
\newcommand{\ee}{\end{equation}}
\newcommand{\xx}{{\mathbf x}}

\newcommand{\qq}{{\mathbf q}}

\begin{document}

\title{Dynamics of long-range ordering in an exciton-polariton condensate}
\author{G.~Nardin$^{1}$}
\author{K.~G.~Lagoudakis$^{1}$}
\author{M.~Wouters$^{2}$}
\author{M.~Richard$^{1,3}$}
\author{A.~Baas$^{1}$}
\author{R. ~Andr\'e$^{3}$}
\author{Le~Si~Dang$^{3}$}
\author{B.~Pietka$^{1*}$}
\author{B.~Deveaud-Pl\'edran$^{1}$}

 \affiliation{$^{1}$ IPEQ, Ecole Polytechnique F\'ed\'erale de Lausanne (EPFL), Station 3, 1015 Lausanne, Switzerland}
 \email[Corresponding author: ]{barbara.pietka@epfl.ch}
\affiliation{$^{2}$ ITP, Ecole Polytechnique F\'ed\'erale de Lausanne (EPFL), Station 3, 1015 Lausanne, Switzerland}
\affiliation{$^{3}$ Institut N\'eel, CNRS, 25 Avenue des Martyrs, 38042 Grenoble, France}

\begin{abstract}
We report on time resolved measurements of the first order spatial coherence in an exciton-polariton Bose-Einstein condensate. Long range spatial coherence is found to set in right at the onset of stimulated scattering, on a picosecond time scale. The coherence reaches its maximum value after the population and decays slower, staying up to a few hundred of picoseconds. This behavior can be qualitatively reproduced, using a stochastic classical field model describing interaction between the polariton condensate and the exciton reservoir within a disordered potential.
\end{abstract}


\maketitle
  
Long range spatial coherence lays at the heart of the remarkable quantum behaviour of matter manifested by superconductors and superfluids at the macroscopic scale. An intriguing question is how quickly the coherent field can be established when a normal incoherent gas is driven through the phase transition. Experimentally, such dynamics are not easily accessible in traditional superconductors and superfluids. Only recently the question of the formation of long range coherence was addressed in experiments with ultracold atomic gases~\cite{ritter-PRL07, hugbart:011602}, while the importance of this study was pointed out even before the realization of Bose Einstein condensation (BEC)~\cite{JETP-92, JETP-94}.

The demonstration of BEC in a solid state system was achieved a few years ago with exciton-polaritons~\cite{Kasprzak-nature}. An exciton-polariton is a quasi-particle that appears in the regime of strong coupling between a microcavity photon and a quantum well exciton. It inherits desirable properties for condensation from both components. Their excitonic content allows for the collisional relaxation, while the photonic component provides a light mass.

Moreover, the polariton statistical properties are directly accessible through the photoluminescence due to a linear relation between external and internal cavity fields~\cite{SavonaPhysRevB}. The cavity radiation field is a part of the polariton wave function. Therefore, the spatial phase coherence, which is the order parameter of a BEC transition~\cite{book-pitaevskii}, can be directly probed. In our experimental work we experimentally measure the first order correlation function in a direct way. Time resolved detection setup in combination with pulsed excitation allows us to observe how the spatial phase correlations are built in time. We therefore probe the time evolution of the first order spatial coherence function~\cite{love:067404, del-valle}: $g^{(1)}(\mathbf{r},\mathbf{r'},\tau=0, t)=<\psi^*(\mathbf{r},t)\psi(\mathbf{r'},t)>$, at zero delay time $\tau$. 

Our sample is the same one as in our previous works~\cite{Kasprzak-nature, Modelocking, richard_PRB05}. It is a CdTe/CdMgTe microcavity grown by molecular beam epitaxy that contains 16 quantum wells with a vacuum field Rabi splitting of 26meV. The sample was mounted in a liquid-He flow cryostat and cooled down to approx. 10K. The sample was excited through a 50x microscope objective of numerical aperture NA = 0.5 providing a diffraction-limited spatial resolution. The excitation spot was tuned to $\sim 15\mu$m diameter in a quasi top-hat intensity profile. We excited the sample non-resonantly with a pulsed Ti:Sapphire laser (250fs pulse duration, $f=80MHz$ repetition rate) tuned to the resonance with energy of the first minimum in the Bragg mirrors at 695nm. In this way we create an electron - hole population above the semiconductor band gap. The photo-created electron - hole pairs first relax into excitons through scattering with optical phonons. These excitons form a thermalized reservoir with a long lifetime of hundreds of ps~\cite{tassonePhysRevB, MullerPhysRevB, bloch:155311}. When the exciton density is sufficiently high, the exciton-exciton scattering rate into the LP branch overcomes the radiative losses and the process of stimulated scattering starts~\cite{bloch:155311}. The polariton population lasts much longer than the polariton life time (few picoseconds). This is possible because the excitonic reservoir keeps replenishing the polariton population.

 \begin{figure}
\begin{center}
\includegraphics[width=65mm]{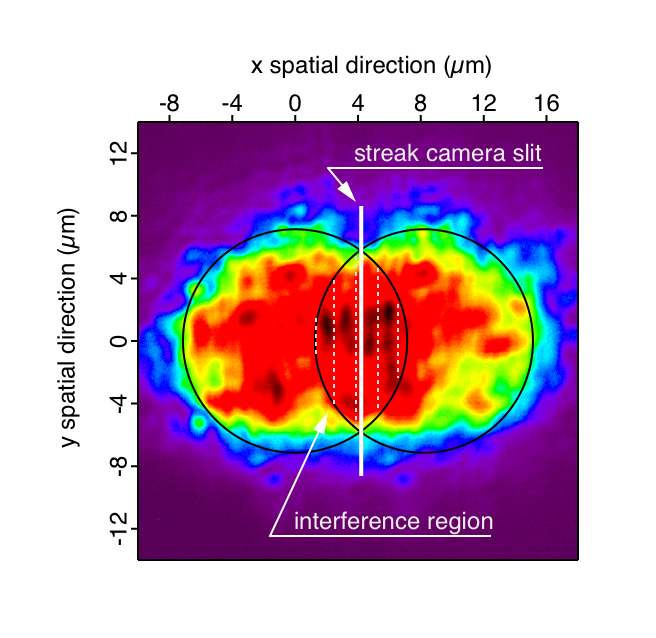} 
\caption{Time integrated real space image of the exciton-polariton condensate emission at the output of the Michelson interferometer. The signal reflected by the two mirrors is marked by black circles. The doted vertical lines are guides to the eye coinciding with the minima of the interference fringes. The part of the photoluminescence in the middle of the overlapped region, at the position of the white solid line, was sent to a streak camera and resolved temporally.}\label{f1}
\end{center}
\end{figure}
\begin{SCfigure*}
\includegraphics[width=130mm]{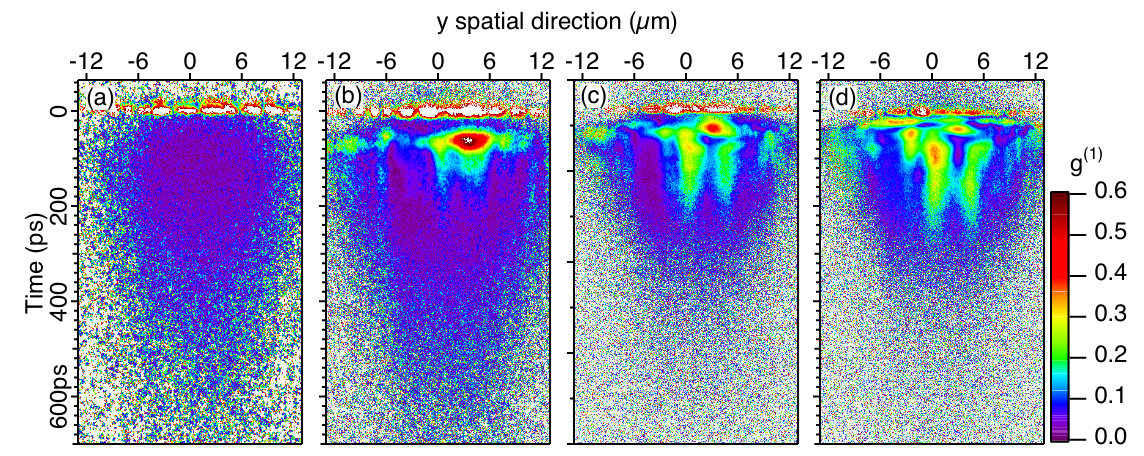}
\caption{Dynamics of first order correlation function, $g^{(1)}$, between two lines of the condensate separated by  8.5$\mu$m for excitation powers (a) 0.5$\cdot P_0$, (b) $P_0$, (c) 1.4$\cdot P_0$ and (d) 2$\cdot P_0$, where $P_0$ is the threshold power. The colour scale denotes the $g^{(1)}$ amplitude, white is used at the positions where the $g^{(1)}$ can not be defined or at the position of the laser pulse at time $t=$0ps.}\label{f2}
\end{SCfigure*}

The coherence of the lower polariton population is probed with a Michelson interferometer in the mirror-mirror configuration. We superimpose the polariton photoluminescence with its copy translated by a distance of $\Delta x$, by tilting one of the mirrors (Fig.~\ref{f1}). In this way, we can investigate the spatial correlations between two points from spatially distant parts of the condensate. Below threshold, the system exhibits short-range order, with a correlation length of $\sim 0.8\mu$m. Choosing the distance $\Delta x=8.5\mu$m, we probe the coherence properties of points separated by more than ten times the thermal de Broglie wavelength. To trace in time the appearance of the condensed phase we have selected a slice in the interference pattern, as it is shown in  Fig.~\ref{f1}, and mapped out its temporal evolution with a streak camera. To relate our temporally resolved interferogram with the first order correlation function we scanned the phase in the Michelson interferometer over $\sim3 \pi$ (see Ref.~\cite{Kasprzak-nature} for more details). 

Fig.~\ref{f2} illustrates the time-resolved images of the buildup of coherence over the selected line in the condensate. The appearance of the long range coherence is a manifestation of the phase transition. If the power in the excitation pulse is not high enough to provide sufficient number of polaritons to cross the condensation threshold, the coherent state is not formed (Fig.~\ref{f2}a). Above threshold, the coherence is formed after a certain delay from the excitation pulse and can persist for more than 100ps. The higher the excitation power, the shorter the delay of the coherence formation. The full dynamics of the coherence (buildup and decay) is strongly accelerated as we increase the excitation power. This acceleration is due to an increasing exciton-exciton collision rate and is an indirect proof of the fact that exciton-exciton collisions play an important role in the exciton relaxation.

\begin{figure}[b]
\begin{center}
\includegraphics[width=86mm]{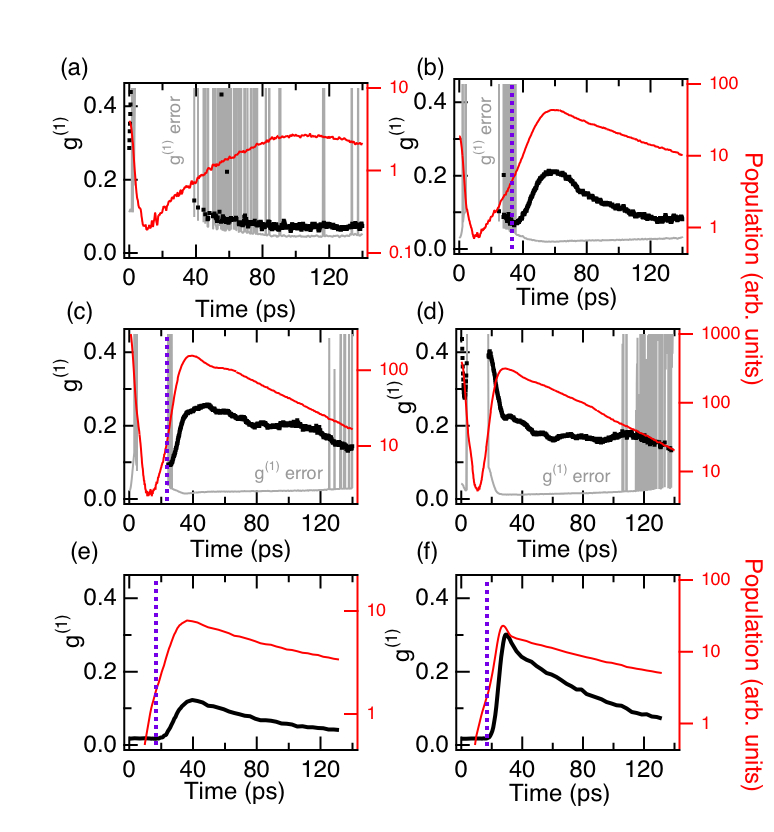} 
\caption{Transient of population and coherence, solid red line and black dots, respectively, for the average over $10\mu m$ along the white solid line in Fig.\ref{f1}, for excitation powers (a) 0.5$\cdot P_0$, (b) $P_0$, (c) 1.4$\cdot P_0$ and (d) 2$\cdot P_0$, where $P_0$ is the threshold power. The grey line shows the estimated standard deviation of the fit coefficients obtained in the $g^{(1)}$ calculation.  (e) and (f) illustrate the theoretical simulation results at and above threshold, respectively. Depicted here is the total polariton population at  $-(\Delta x/2)$ and $(\Delta x/2)$ positions. The onset of the stimulated scattering is marked by the vertical dashed lines.}\label{f3}
\end{center}
\end{figure}

The inhomogeneous distribution of photoluminescence intensity within the excitation spot (Fig.~\ref{f1}) and the coherence in the spatial direction results from the disorder in the sample that modulates the polariton density~\cite{Modelocking}. The time evolution of coherence shows also an irregular behaviour. The higher the excitation power the more the coherence fluctuates in time. This effect is due to redistribution of polaritons over the excitation spot as the effective disorder seen by the particles is changing with the population~\cite{Modelocking}. To get a general understanding of the coherence behaviour, less influenced by the disorder, we have averaged the observed coherence over the different correlated points, that always have the same spatial separation of $\Delta x=8.5\mu$m between them (see Fig.~\ref{f3}).

The time evolution of the polariton density is traced (Fig.~\ref{f3} red line) simultaneously with the coherence (Fig.~\ref{f3} black dots), by temporally resolving the emission reflected separately by the two mirrors of the Michelson interferometer. This gives a more detailed insight in the time at which the phase coherence starts to appear with respect to the particle population. The dashed vertical lines in Fig.~\ref{f3} separate the low  polariton density region, from the region of high density. At this border the density evolution shows a cusp, indicating the onset of stimulated scattering. Exactly at the same time (below our temporal resolution limit of 3.5ps), the coherence starts to build up. This directly shows that the communication between points separated by $8.5\mu$m in the condensate so as to define a common phase, is very fast, in a picosecond time scale. It gives a phase communication velocity faster than $2.5\cdot 10^6$m/s. The total time span of the condensate formation can be evaluated by the time needed for an increase of the coherence from 10\% to 90\% of the maximal value. We thus obtain 12ps at the threshold $P_0$, and 6.5ps for an excitation power of 1.4$\cdot P_0$. This means that the total coherence is established here with a velocity $0.7 \cdot 10^6$m/s at threshold and $1.3 \cdot 10^6$m/s for 1.4$\cdot P_0$. The higher the excitation power the faster the coherent phase is built. These velocities are faster or comparable with the velocity of sound in the polariton BEC, $c=\sqrt{\mu/m}$. We estimate $c$ from the blue shift $\mu$ induced by the condensate polaritons and the lower polariton mass $m$, to be $(0.7\pm0.2)\cdot 10^6$m/s. This means that the appearance of coherent phase is faster than any process mediated by the interactions. Our result differs in this case from the cold atomic gases~\cite{ritter-PRL07}, where it was found that the coherent region grows with a velocity of approximately a factor of five slower than the speed of sound in atomic BEC. Atomic BEC are closed systems, therefore the only way for correlations to propagate is using interactions within the atomic cloud. Polaritons form a driven dissipative system. Moreover, the condensed fraction of polaritons is in constant interactions with the excitonic reservoir and phonons. Therefore the correlations can be indeed established faster than the speed of sound. Once it is established, the phase coherence slowly decays, and remarkably enough, it is maintained in the system for a long time, of about one hundred ps, decaying slower than the population for high excitation powers. 

\begin{figure}
\begin{center}
\includegraphics[width=86mm]{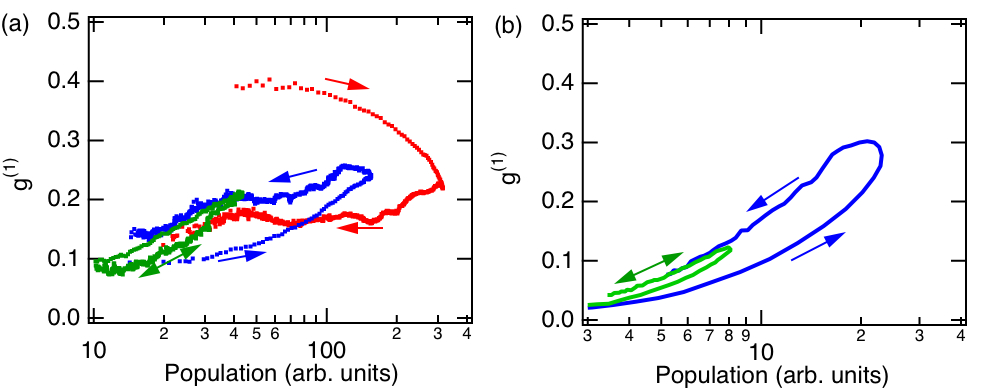}
\caption{(a) Experimental and (b) theoretical results of build up and decay of coherence as the population of polaritons is formed at threshold (green), above threshold (blue) and twice above threshold (red). The timeline is marked by arrows. For high excitation powers we evidence the appearance of a hysteresis between the rise and decay.}\label{f4}
\end{center}
\end{figure}

A comparison of the data measured at the different pump powers can be made in Fig.~\ref{f4}, where the coherence is plotted versus the density for the three different pump intensities. This figure shows the importance of the speed at which the phase transition is crossed. At low pump intensity (green curve), where the density builds up slowly, the coherence is almost a single valued function of the condensate density. Physically, this can be interpreted as an adiabatic evolution of the coherence with the density.
At higher powers (blue curve) on the other hand, a hysteresis loop opens up. At initial times, the coherence is below the steady state value. During the fast density ramp, the phase coherence does not have time to spread through the system.

The same qualitative behaviour is obtained in theoretical simulations~\cite{wouters_prb09}. Our model describes the lower polariton dynamics by a stochastic classical field equation coupled to a rate equation for the exciton reservoir. The latter is split in two parts: the active excitons that can match energy and momentum conservation for a scattering process into the lower polariton branch (density $n_{R2}$) and the excitons that cannot directly scatter into the lower polariton branch (density $n_{R1}$). The coupled equations of motion of the reservoirs describe the nonresonant excitation with intensity $P$, the equilibration of the densities between the two reservoirs at a rate $\kappa_{12}$ and the scattering into the lower polariton region
\begin{eqnarray}
\frac{d}{dt} 
n_{R1}&=&P-\kappa_{12}(\frac{n_{R1}}{r_{12}}-n_{R2})-\gamma_{R1}n_{R1} \\
\frac{d}{dt} n_{R2}&=& 
\kappa_{12}(\frac{n_{R1}}{r_{12}}-n_{R2})-\gamma_{R2}n_{R2} 
-\alpha \left.\frac{d n_{R 2}}{dt}\right\vert_{R2\rightarrow LP},
\label{eq:motres12}
\end{eqnarray}
where $r_{12}$ is the equilibrium ratio between the two excitonic 
reservoirs and the last term on the second line accounts for the scattering into the lower polariton region. The coupling terms between the reservoir and lower polariton 
branch read explicitely $\mathcal R_{\rm in,out} \psi(\xx) =  \sum_{\qq} e^{-i\qq \xx'} \sqrt{n_{R2}(\xx)n_{R2}(\xx')} R_{\rm in, out}(\epsilon_\qq)  \psi_{\xx'} $
The second reservoir density is normalized so that threshold is reached 
when it is unity. The parameter $\alpha$ quantifies the depletion of the 
exciton reservoir due to its scattering into the lower polariton field 
$\psi$, defined on a spatial grid with unit cell area $\Delta V$, that 
is described by the stochastic classical field equation
\begin{eqnarray*}
i d \psi =  dt \left [\frac{-\hbar^2 \nabla^2}{2m}+ \frac{i(\mathcal 
R_{\rm in}-\mathcal R_{\rm out}-\gamma)}{2}+ V_{\rm ext}
g |\psi(\xx)|^2 \right]\psi \\
+ dW,
\label{eq:stoch_mot}
\end{eqnarray*}
where $V_{\rm ext}$ is the disorder potential that acts on the polaritons. The noise term $dW$ is a complex Gaussian stochastic variable with the 
correlation functions proportional to the sum of loss and gain.
The advantage of this method for the modeling of the experiments is that spatially inhomogeneous systems can be described without any additional cost. We are thus able to model a disordered microcavity excited with a spatially finite laser pulse.

Figs.~\ref{f3}e, f show the results for two different excitation intensities~\cite{text2}. They show qualitatively the same trend as the experimental data. Close to threshold (Fig.~\ref{f3}e), the coherence follows the population, while it stays behind further above threshold (Fig.~\ref{f3}f), and is kept for longer times. In the coherence versus population graph (Fig.~\ref{f4}b), the hysteresis loops appear (blue curve) and as in the experimental data its upper part overlaps with the curve at lower pump intensity (green curve). 

The disorder potential limits the coherence of the polariton condensate. Simulations without disorder predict a maximal coherence of about 0.45. Less expected is that the finite size excitation spot also plays a role in the coherence formation time. With respect to an infinite system (simulated by using periodic boundary conditions), the coherence builds up faster in the finite system. The physical reason is that spontaneously formed vortices in the condensate phase~\cite{kibble, zurek, spontan_vortices} can escape away from the system at its boundaries, instead of only annihilating each other.
Finally, simulations have shown that the out-scattering term in the model (see equation (\ref{eq:stoch_mot})) plays an important twofold role. On one hand, it decreases the time needed to build up the coherence, on the other hand, it decreases the maximal value of the coherence that is achieved \cite{wouters_prb09}. The latter effect is due to the enhancement of fluctuations, whereas the initial acceleration of the coherence formation is due to the out-scattering contribution to the deterministic term that quickly eliminates the high momentum components of the polariton gas. Conceptually, this is important, because it shows that the coherence of the low energy polaritons is strongly influenced by their interaction with the excitonic reservoir. The polariton subsystem behaves in this way differently from ultracold atomic gases, which are isolated from their environment. 

Let us now comment on the experimental data at the highest pump intensity (Fig.~\ref{f3}d and the red curve in Fig.~\ref{f4}a). In this ultrafast regime, the coherence builds up surprisingly fast. Not only does the hysteris loop (Fig.~\ref{f4}a red curve) turn in the other direction with respect to the blue curve, but the coherence at short times is even above the value that is reached in the adiabatic regime. Within our theoretical model, we have not been able to find a regime where this behaviour is realized. It is possible that it originates from a nontrivial dynamics in the exciton reservoir, but could also be related to the finite time measurements. Note that the error on the coherence at short times (gray line in Fig.~\ref{f3}d) is large at the initial time where the coherence is formed. 

In conclusion, our work has provided substantially new experimental observations on the formation dynamics of the long range order in the exciton-polariton BEC. The all-optical access of the polariton condensate properties is a clear advantage to study the dynamics of its coherence formation. We have demonstrated that the onset of phase coherence between points separated by $8.5\mu$m in the condensate appears in a picosecond time scale, much faster than expected: simultaneously with the onset of stimulated scattering. When the dynamics in the system is slow, at threshold, the coherence adiabatically follows the particle population. When the system is rapidly driven through the phase transition, the maximum long-range order is established in the system slower than the buildup of the population. This demonstrates that the system needs some time, a few picoseconds in our case, to buildup the highest long range order. Once it is formed, the coherence can be maintained in the polariton system for more than a hundred picoseconds. We have found that the velocity of the establishment of the phase in a polariton condensate is as fast or faster than the sound velocity, i.e. much faster than in cold atom systems. These experimental results are supported by a theoretical model which brings new elements to the understanding of the mechanisms that regulate the phase mediation in non-equilibrium condensates. The observation and the understanding of these ultrafast dynamics are also of crucial importance for the design of future polariton laser devices.

The work was supported by the Swiss National Research Foundation Through NCCR "Quantum Photonics".



\begin{thebibliography}{10}
\bibitem{text2}
In the reported simulations, the parameters were taken 
$m_{LP}/\hbar^2=1.5 \mu m^{-2} meV^{-1} $, 
$\hbar \gamma = 0.5 meV$,
$g=10^{-3} meV \mu m^{2}$,
$k_B T_R=2 meV$,
$R_{out}(\tilde n_R,\epsilon)=6 \tilde n_R meV$,
$R_{in}(\tilde n_R,\epsilon)=
(\gamma+6 meV)\exp(-\epsilon/k_B T_R)\tilde n_R $,
$\alpha=0.01$,
$\kappa_{12}=0.25$,
$r_{12}=30$,
$\hbar \gamma_{R1}=\hbar \gamma_{R2}=1\mu eV$,
and a disorder potential with Gaussian distribution of the amplitude with a variance of 0.8meV and a correlation length of 4$\mu$m. A 10ps laser pulse was applied to excite the reservoir excitons. Numerical results were obtained by Monte Carlo averaging over 250 realizations over the noise.
\end{thebibliography}
\end{document}